\documentclass[12pt]{article}
\usepackage{amsmath,amssymb}
\usepackage{graphicx}
\usepackage{amsmath}
\usepackage{hyperref}
\usepackage{float}
\newcommand {\be}{\begin{equation}}
 \newcommand {\ee}{\end{equation}}
 \newcommand {\bea}{\begin{array}}
 
 \newcommand {\eea}{\end{array}}

\textwidth=6.5in \hoffset=-.75in \voffset=-1in \numberwithin{equation}{section}
\numberwithin{figure}{section}

\def\0{{(0)}}
\def\1{{(1)}}
\def\2{{(2)}}

\def\<{\langle }
\def\>{\rangle }
\def\[{\left[}
\def\]{\right]}

\begin{document}
\begin{titlepage}

\vskip1cm
\begin{center}
{~\\[140pt]{ \LARGE {\textsc{Holographic Entanglement Entropy in flat limit of the Generalized Minimal Massive Gravity model }}}\\[-20pt]}
\vskip2cm

\end{center}
\begin{center}
{M. R. Setare \footnote{E-mail: rezakord@ipm.ir}\hspace{1mm} ,
M. Koohgard \footnote{E-mail: m.koohgard@modares.ac.ir}\hspace{1.5mm} \\
{\small {\em  {Department of Science,\\
 Campus of Bijar, University of Kurdistan, Bijar, Iran }}}}\\
\end{center}
\begin{abstract}
Previously we have studied the Generalized Minimal Massive Gravity
(GMMG) in asymptotically $AdS_3$ background, and have shown that the theory is free of negative-energy bulk
modes. Also we have shown GMMG avoids the aforementioned “bulk-boundary unitarity
clash”. Here instead of $AdS_3$ space we consider asymptotically flat space, and study this model in the flat limit. The dual field theory of GMMG in the flat limit is a $BMS_3$ invariant
field theory, dubbed (BMSFT) and we have BMS algebra asymptotically instead of Virasoro algebra. In fact here we present an evidence for this claim. Entanglement entropy of GMMG is calculated in the background in the flat null infinity. Our evidence for mentioned claim is the result for entanglement entropy in filed theory side and in the bulk (in the gravity side). At first using Cardy formula and Rindler transformation, we calculate
entanglement entropy of BMSFT in three different cases.  Zero
temperature on the plane and on the cylinder, and non-zero temperature case. Then we obtain the entanglement entropy in the bulk. Our results in gravity side are exactly in agreement with field theory calculations.

\end{abstract}
\vspace{1cm}

Keywords: Entanglement Entropy, Cardy formula, GMMG model, BTZ black hole

\end{titlepage}

\section{Introduction}	
Holography \cite{1,2} is a common property for quantum gravity that relates a gravity theory living in a bulk to a quantum field theory living in its boundary. A good sign for holography is the entanglement entropy, which is a good description of the correlation of the theory on the bulk and the CFT theory on the boundary. Entanglement entropy calculation can be done in 2D conformal field theories because infinite-dimensional Virasoro algebra, which generates 2D conformal transformations, can be used to simplify calculations \cite{3,4,5}. For this purpose, the entropy is calculated for an regularized interval to avoid ultra-violet divergences. The role of holography is to create an equivalence between a gravity theory and a quantum field theory without gravity at its boundary. The result of the holography is that it is conjuctured that entanglement entropy of the conformal field theory is equivalent to the calculation of the area of an extremal codimension-two-surface in AdS \cite{6}.

Rindler transformation method \cite{7} maps entanglement entropy to thermal entropy with the help of symmetry transformations. In this method, the Bekenestein-Hawking entropy calculates the thermal entropy and the thermal entropy is equal to the entanglement entropy. This is because on the boundary and on the bulk, respectively, with the help of some transformations, we map the entanglement entropy to the thermal entropy on a hyperbolic spacetime and map the AdS vacuum to black holes with a hyperbolic horizon. Then, with the help of the AdS/CFT dictionary \cite{8,9,10}, we can calculate the thermal entropy using the Bekenestein-Hawking entropy and obtain the entanglement entropy.

In recent years, some interesting models of massive gravity in three dimensions have been
introduced. Among these models, a well known one is that of the generalized minimal massive gravity (GMMG) \cite{11}. This model is realized by adding the CS deformation term, the higher derivative
deformation term, and an extra term to pure Einstein gravity with a
negative cosmological constant. Usually the theories including the terms given by the square of the curvatures have the massive spin
2 mode and the massive scalar mode in addition to the massless graviton. Also the theory has ghosts
due to negative energy excitations of the massive tensor. In \cite{11} it is discussed that the GMMG is free
of negative-energy bulk modes, and also avoids the aforementioned
\emph{bulk-boundary unitarity clash}. Such clash is present in the previously constructed
gravity theories with local degrees of freedom in 2+1-dimensions, namely Topologically
Massive Gravity \cite{11', 11"} and the cosmological extension of New Massive Gravity \cite{12'}. By a Hamiltonian
analysis one can show that the GMMG model
has no Boulware-Deser ghosts and this model propagate
only two physical modes. Since this theory avoid the bulk-boundary clash, it define excellent arenas to explore
the structure of asymptotically AdS solutions, asymptotic symmetries, its algebra and other
holographically inspired questions. In this paper we study a holographic calculation of entanglement entropy in asymptotically flat geometry solutions of this model.

It is well known that the group of asymptotic symmetries of asymptotically flat spacetimes at future null infinity is the BMS group \cite{12,13,14}.
In extension of AdS/CFT correspondence to the flat space holography, the BMS algebra has been investigated very much in recent years \cite{15,16,17,18,19,20,21,22}. In this paper, we use the BMS gauge to do the computations. The entanglement is also considered to be proportional to the length of the spacelike geodesic, which is one of the three geodesics characteristic of three-dimensional flat spacetime.

The paper is organized as follows. In the section [2] we will focus on the field theory with the BMS symmetry, and with the help of Cardy formula and Rindler transformaation, we calculate entanglement entropy of GMMG model. In the section [3], the spacetime metric is described in the gravity side and the flat spacetime is divided into three classes. The entanglemnet entropy of the gravity side is then calculated, in which we consider the entropy proportional to the spacelike geodesic length. We present some conclusions in section [4].
	\section{Entanglement entropy in field theory side}
	It has already been shown that the GMMG model, if written in the background of the $AdS_3$ space, is dual with a two-dimensional conformal field theory, at least in part of the space of the parameters of the dual model. In this paper, we study this model in flat limit and we provide an important proof of this correspondence and duality, and that is the calculation of the entanglement entropy on both sides of the duality. We know that the model, when re-written in flat limit and in the background of an asymptotically flat spacetime, is dual with a field theory with the $BMS_{3}$ symmetry called the BMSFT. We look at the field theory in three different cases. Using the Cardy formula and calculations that have already been obtained for entanglement entropy  of these cases of the field theories, we find the entanglement entropy of the field theory. Then we find the general form of the entanglement entropy in asymptotically flat limit and we consider the duality in the three cases. By finding the entanglement entropy, we show that the result on each of the cases on both sides of the duality is exactly the same. In this way, we obtain a strong and reasoned proof of the gauge/gravity duality.
The symmetry of BMS is determined by transformations whose the charge algebra is as follows \cite{23}
\begin{eqnarray}
  \big[\mathcal{L}_{n},\mathcal{L}_{m}\big] &=& (n-m)\mathcal{L}_{n+m}+\frac{c_L}{12}(n^3-n)\delta_{m+n,0}\nonumber \\
  \big[\mathcal{L}_n,\mathcal{M}_m\big] &=& (n-m)\mathcal{M}_{n+m}+\frac{c_M}{12}(n^3-n)\delta_{m+n,0}\nonumber \\
  \big[\mathcal{M}_n,\mathcal{M}_m\big] &=& 0
\end{eqnarray}
where $c_L$ and $c_M$ are the central charges and $\mathcal{L}_n$ and $\mathcal{M}_n$ are the conserved charges of the currents that generate the following $BMS$-transformations on the boundary coordinates $(u,\phi)$
\begin{eqnarray}\label{first tr}
  \tilde{\phi} &=& f(\phi)\nonumber \\
  \tilde{u} &=& \partial_{\phi}f(\phi)u+g(\phi),
\end{eqnarray}
where $f(\phi)$ and $g(\phi)$ are two arbitrary functions.

In the Bondi gauge, $2+1$-dimensional solution of the Einstein's equation in the flat plane is as follows \cite{20}

\begin{equation}\label{general met 2}
  ds^2=M du^2-2 dudr+ J du d\phi+r^2d\phi^2
\end{equation}
in which the cosmological constant is considered equal to zero. The flat metric Eq.~(\ref{general met 2}) is categorized in three cases based on the $J$ and $M$ values. When $M=-1$ and $J=0$, the metric Eq.~(\ref{general met 2}) is converted to the Minkowski metric in three dimensions (called global Minkowski solution), as follows
\begin{equation}\label{general met 222}
  ds^2= - du^2-2 dudr+r^2d\phi^2
\end{equation}

The second case is for the solutions that $M=J=0$, which shows a manifold called the null orbifold. This orbifold is defined by a null boost and was first used in the string theory \cite{26}. This metric in this case has the following form
\begin{equation}\label{p met}
  ds^2= -2 dudr+r^2d\phi^2
\end{equation}

The third case is considered with the $M>0$ in which we have a flat space cosmological geometry (FSC). This is a time-dependent solution of Einstein equation in three dimensions of spacetime \cite{27}. The flat space metric Eq.~(\ref{general met 2}) has an Cauchy event horizon that is best seen in the ADM form as follows
\begin{equation}\label{flat met in ADM}
  ds^2=-(\frac{J^2}{4r^2}-M)^2dt^2+(\frac{J^2}{4r^2}-M)^{-1}dr^2+r^2(d\phi+\frac{J}{2r^2}dt)^2
\end{equation}
where
\begin{equation}\label{Cauchy h}
  r_H=\mid r_c\mid\equiv \mid \frac{J}{2\sqrt{M}}\mid
\end{equation}
is location of the horizon. We consider the thermal circle of the FSC metric Eq.~(\ref{general met 2}) by defining the parameters as follows
\begin{equation}\label{thermal circle}
  (u,\phi)\sim (u+i\beta_u,\phi-i\beta_{\phi});~~~\beta_u=\frac{\pi J}{M^{3/2}},~~~\beta_{\phi}=\frac{2\pi}{\sqrt{M}}
\end{equation}

In the  Rindler transformation, the symmetry translations are generated along the thermal circle. This circle is produced by some imaginary identifications between coordinates. \\
The second metric case mentioned above is very similar to the Poincare patch in the $AdS_3$ spacetime, which we can easily see by the following transformation between the Poincare coordinates and the Cartesian coordinates.
\begin{eqnarray}
  t &=& \frac{l_{\phi}}{r}+\frac{2}{l_{\phi}}(u+\frac{r\phi^2}{2}), \\
  x &=& \frac{l_u}{l_{\phi}}+r\phi, \\
  y &=& \frac{l_{\phi}}{r}-\frac{2}{l_{\phi}}(u+\frac{r\phi^2}{2}).
\end{eqnarray}

Using the coordinate transformation in Eq.~(\ref{p met}),  we have the metric form as a Poincare spacetime as follows
\begin{equation}\label{p met 2}
  ds^2=-2 dudr+r^2d\phi^2=-dt^2+dx^2+dy^2
\end{equation}

Since there are three cases for the background spacetime Eq.~(\ref{general met 2}), we consider the field theory in three cases on the boundary: zero temperature on the plane and on the cylinder, and non-zero temperature. The periodicity along imaginary time axis characterizes temperature of a quantum field theory. By considering imaginary time, the thermal circle changes to spatial circle.

 We want to find a Rindler transformation that relates the entanglement entropy to the thermal entropy. The thermal entropy of BMSFT is calculated on the following regularized interval
\begin{equation}\label{reg. int}
  (-l_u/2+\epsilon_u,~-l_{\phi}/2+\epsilon_{\phi})~\to~(l_u/2-\epsilon_u,~l_{\phi}/2-\epsilon_{\phi})
\end{equation}
where $l_u$ and $l_{\phi}$ are the interval extensions along $u$ and $\phi$ directions. The Cardy formula is used to calculate the entropy of two-dimensional CFTs, and with this formula, the thermal entropy is obtained as an asymptotic density of states in a statistical manner \cite{29}. To calculate the thermal entropy of the field theory, we use following form of Cardy formula
\begin{equation}\label{Cardy formula}
  S_{\bar{b}|b}(\bar{a}|a)=-\frac{\pi^2}{3}\bigg(c_L\frac{b}{a}+c_M\frac{(\bar{a}b-a\bar{b})}{a^2}\bigg)
\end{equation}
where $(\bar{a}, a)$ specifies the thermal circle and $(\bar{b}, b)$ specifies the spatial circle, which we apply the coordinate changes in the field theory using the Rindler transformation between these two circles. The Cardy formula is obtained by counting the states that identify the spatial circle and the thermal circle; the manifold of the BMSFT is defined on the following torus
\begin{equation}\label{BMSFT torus}
  (\tilde{u}, \tilde{\phi})\sim(\tilde{u}+i\bar{a},~\tilde{\phi}-i a)\sim(\tilde{u}+2\pi\bar{b},~\tilde{\phi}-2\pi b)
\end{equation}

The coordinate identification in the form $\tilde{x}^i\sim \tilde{x}^i+i\tilde{\beta}^i$ is called the thermal identification. The second identification in Eq.~(\ref{BMSFT torus}) is a spatial circle that preserve the metric form as follows
\begin{equation}\label{metric by spatial cir}
  ds^2=\tilde{M}d\tilde{u}^2-2d\tilde{u}d\tilde{r}+\tilde{J}d\tilde{u}d\tilde{\phi}+r^2d\tilde{\phi}^2
\end{equation}

The coordinates used are the Bondi coordinates. The BMSFT is invariant under the Rindler transformations, and at the same time these transformations preserve $BMS_{3}$ symmetry of the theory asymptotically. Rindler transformation as a coordinate transformation is as follows \cite{27'}
\begin{eqnarray}\label{rindler}
  \tilde{\phi} &=& f(\phi)\nonumber \\
  \tilde{u} &=& \partial_{\phi}f(\phi)u+g(\phi)
\end{eqnarray}
where its generators are considered as a combination of BMS generators as follows
\begin{eqnarray}\label{rindler gen}
  \partial_{\tilde{\phi}} &=& \Sigma_{n=-1}^1(b_n\mathcal{L}_n+d_n\mathcal{M}_n)\nonumber \\
  \partial_{\tilde{u}} &=& -\Sigma_{n=-1}^1(b_n\mathcal{M}_n)
\end{eqnarray}
where $\mathcal{L}_n$ and $\mathcal{M}_n$ are $BMS_{3}$ generators as follows \cite{23}
\begin{eqnarray}\label{BMS gens}
  \mathcal{L}_n &=& -u(n+1)\phi^n\partial_u-\phi^{n+1}\partial_{\phi},\nonumber \\
  \mathcal{M}_n &=& \phi^{n+1}\partial_u
\end{eqnarray}

By Rindler generators and by $k_t\equiv\tilde{\beta}^i\partial_{\tilde{x}^i}$, we can create a translation in the direction of the thermal circle as $\tilde{x}^i(s)=\tilde{x}^i+\tilde{\beta}^is$, and in general, an identification is considered as $\tilde{x}^i\sim \tilde{x}^i(i)$. These transformations are generally obtained on the plane as follows \cite{24}
\begin{equation}\label{rindler explicit1}
  \tilde{\phi}= \frac{\tilde{\beta}_{\phi}}{\pi}\tanh^{-1}(\frac{2\phi}{l_{\phi}})
\end{equation}

\begin{equation}\label{rindler explicit2}
  \tilde{u}+\frac{\tilde{\beta}_u}{\tilde{\beta}_{\phi}}\tilde{\phi} = \frac{2\tilde{\beta}_{\phi}(u l_{\phi}-l_u\phi)}{\pi(l^2_{\phi}-4\phi^2)}
\end{equation}

By substituting the beginning and the end of the regularized interval Eq.~(\ref{reg. int}) into Eqs. ~(\ref{rindler explicit1})~and~(\ref{rindler explicit2}), we have the interval length as follows
\begin{eqnarray}\label{rindler first}
  \Delta\tilde{\phi} &=& \frac{\tilde{\beta}_{\phi}}{\pi}\big(\tanh^{-1}(\frac{2\phi_{max}}{l_{\phi}})
  -\tanh^{-1}(\frac{2\phi_{min}}{l_{\phi}})\big)\nonumber \\
   &=& \frac{\tilde{\beta}_{\phi}}{2\pi}\big(\log(\frac{l_{\phi}+2\phi_{max}}{l_{\phi}-2\phi_{max}})  -\log(\frac{l_{\phi}+2\phi_{min}}{l_{\phi}-2\phi_{min}})\big)\nonumber \\
   &=& \frac{\tilde{\beta}_{\phi}}{2\pi}\big(\log(\frac{2 l_{\phi}-2\epsilon_{\phi}}{-2\epsilon_{\phi}})-\log(\frac{2\epsilon_{\phi}}{2 l_{\phi}-2\epsilon_{\phi}})\big)\nonumber \\
   & \simeq & \frac{\tilde{\beta}_{\phi}}{\pi}\log(\frac{l_{\phi}}{\epsilon_{\phi}})
\end{eqnarray}
and
\begin{eqnarray}\label{rindler 2nd}
  \Delta\tilde{u} &=& \frac{2\tilde{\beta}_{\phi}(u_{max}l_{\phi}-\phi_{max}l_u)}{\pi(l_{\phi}-2
  \phi_{max})(l_{\phi}+2\phi_{max})}-\frac{2\tilde{\beta}_{\phi}(u_{min}l_{\phi}-\phi_{min}l_u)}{\pi(l_{\phi}-2
  \phi_{min})(l_{\phi}+2\phi_{min})}-\Delta\tilde{\phi} \nonumber\\
   &=& \frac{\tilde{\beta}_{\phi}}{\pi}\frac{\epsilon_{\phi}l_u-\epsilon_ul_{\phi}}{\epsilon_{\phi}
   (l_{\phi}-\epsilon_{\phi})}-\Delta\tilde{\phi} \nonumber\\
   & \simeq & \frac{\tilde{\beta}_{\phi}}{\pi}
   (\frac{l_u}{l_{\phi}}-\frac{\epsilon_u}{\epsilon_{\phi}}-\log(\frac{l_{\phi}}{\epsilon_{\phi}}))
\end{eqnarray}

Therefore, the interval of the coordinate extensions on the plane is as follows
\begin{equation}\label{F. int}
  \mathcal{I}_{reg}:~~~\big(-\frac{\Delta\tilde{u}}{2}, -\frac{\Delta\tilde{\phi}}{2}\big)~\to~
  \big(\frac{\Delta\tilde{u}}{2}, \frac{\Delta\tilde{\phi}}{2}\big)
\end{equation}

We use the regularized interval Eq.~(\ref{F. int}) for the BMSFT with zero temperature on the plane. Here we consider the $\tilde{\beta}_{\phi}$ and the $\tilde{\beta}_u $ to determine the thermal circle (i.e. $ a $ and $\tilde{a}$) as follows
\begin{equation}\label{a, beta}
  a=\tilde{\beta}_{\phi},~~~~\bar{a}=\tilde{\beta}_u
\end{equation}

The extensions $\Delta\tilde{u}$ and $-\Delta\tilde{\phi}$ are considered corresponding to $ 2\pi\bar{b}$ and $ 2\pi b $ respectively to determine the spatial circle, as follows
\begin{equation}\label{b, delta}
  2\pi b=-\Delta\tilde{\phi}, ~~~~2\pi\bar{b}=\Delta\tilde{u}
\end{equation}

On the other hand, the field theory is discussed on a canonical torus, which is characterized by $\phi\sim\phi+2\pi $. Using the BMS transformation as follows
\begin{equation}\label{can torus}
  \hat{\phi}=\frac{\tilde{\phi}}{b}, ~~~~~\hat{u}=\frac{\tilde{u}}{b}+\frac{\bar{b}}{b^2}\tilde{\phi}
\end{equation}
manifold of the field theory is characterized by the following identifications
\begin{eqnarray}\label{Identi..s}
  (\hat{u},\hat{\phi}) &\sim & (\hat{u}+i \hat{\beta}_u, \hat{\phi}+i \hat{\beta}_{\phi})\sim (\hat{u},\hat{\phi}-2\pi)\nonumber \\
  \hat{\beta}_{\phi} &=& \frac{a}{b}~~~~\hat{\beta}_{u}=\frac{\bar{a}b-a\bar{b}}{b^2}
\end{eqnarray}

Thus, according to the regularized interval extensions Eqs.~(\ref{rindler first}) and~(\ref{rindler 2nd}), we have the following relations
\begin{eqnarray}\label{beta hat}
  \hat{\beta}_{\phi} &=& -\frac{2\pi^2}{\log(\frac{l_{\phi}}{\epsilon_{\phi}})}\nonumber \\
  \hat{\beta}_u &=& -\frac{\hat{\beta}_{\phi}^2}{2\pi^2}(\frac{l_u}{l_{\phi}}-\frac{\epsilon_u}{\epsilon_{\phi}})
\end{eqnarray}
where we have used the identifications Eqs.~(\ref{a, beta}) and~(\ref{b, delta}). By substituting the 2nd line of (\ref{Identi..s}) into the Cardy formula~(\ref{Cardy formula}) for the thermal entropy , we have the following result
\begin{equation}\label{Card-mid}
  S=-\frac{\pi^2}{3}\bigg(c_L\frac{1}{\hat{\beta}_{\phi}}+c_M\frac{\hat{\beta}_u}{\hat{\beta}_{\phi}^2}\bigg)
\end{equation}
where by substituting the results ~(\ref{beta hat}) into this entropy formula, we have the general form of the entanglement entropy for BMSFT with zero temperature on the plane, as follows

\begin{equation}\label{EE for 0 plane}
  S_{EE}=\frac{c_L}{6}\log(\frac{l_{\phi}}{\epsilon_{\phi}})
  +\frac{c_M}{6}(\frac{l_u}{l_{\phi}}-\frac{\epsilon_u}{\epsilon_{\phi}})
\end{equation}

For the flat limit of GMMG where the cosmological constant is considered to be zero, the central charges are as follows \cite{25}
\begin{eqnarray} \label{cent ch.}
  c_L &=& -\frac{3}{\mu G}\nonumber \\
  c_M &=& -\frac{3}{G}(\bar{\sigma}+\frac{\alpha H}{\mu}+\frac{F}{m^2})
\end{eqnarray}

By substituting the central charges Eq.~(\ref{cent ch.}) in entanglement entropy Eq.~(\ref{EE for 0 plane}), the entanglement entropy for the zero temperature BMSFT on the plane is as follows
\begin{eqnarray} \label{EE in zero BMSFT}
  S_{EE} &=& -\frac{1}{2\mu G}\log(\frac{l_{\phi}}{\epsilon_{\phi}})-\frac{1}{2 G}
  (\bar{\sigma}+\frac{\alpha H}{\mu}+\frac{F}{m^2})(\frac{l_u}{l_{\phi}}-\frac{\epsilon_u}{\epsilon_{\phi}}) \nonumber\\
   &=& -\frac{1}{4 G}(\bar{\sigma}+\frac{\alpha H}{\mu}+\frac{F}{m^2})(\sqrt{\tilde{M}}\Delta\tilde{u}+
   \frac{\tilde{J}}{2\sqrt{\tilde{M}}}\Delta\tilde{\phi})-\frac{1}{4\mu G}\sqrt{\tilde{M}}\Delta\tilde{\phi}
\end{eqnarray}

In the last line of this equation we have used Eqs.~(\ref{rindler first})~and~(\ref{rindler 2nd}). We do the same calculations for other two cases of the BMSFTs. In this case we have the following identification
\begin{equation}\label{finite ident.}
  (u,\phi)\sim (u+i\beta_u,\phi-i\beta_{\phi})
\end{equation}

By similar steps as in the zero temperature BMSFT on the plane, parameters of the Cardy formula are obtained based on Rindler transformation as follows \cite{24}.
\begin{eqnarray} \label{rindler002}
  a &=& \tilde{\beta}_{\phi},~~~~ \bar{a}=\tilde{\beta}_u \nonumber \\
  2\pi b &=& - \frac{\tilde{\beta}_{\phi}}{\pi}\zeta,~~~~ 2\pi\bar{b}=-\frac{\tilde{\beta}_{u}}{\pi}\zeta+
  \frac{\tilde{\beta}_{\phi}}{\pi \beta_{\phi}}\big[\pi\big(l_u+\frac{\beta_u}{\beta_{\phi}}\big)
  \coth\frac{\pi l_{\phi}}{\beta_{\phi}}-\beta_u\big],
\end{eqnarray}
where $\zeta=log\big(\frac{2}{\epsilon_{\phi}}\sin \frac{l_{\phi}}{2}\big)$. Substituting these parameters into the Cardy formula (\ref{Cardy formula}), the entanglement entropy of the BMSFT with finite temperature is as follows
\begin{equation}\label{EE in Poin}
  S_{EE}=\frac{c_L}{6}\log\big(\frac{\beta_{\phi}}{\pi\epsilon_{\phi}}\sinh\frac{\pi l_{\phi}}{\beta_{\phi}}\big)+\frac{c_M}{6}\frac{1}{\beta_{\phi}}\big[\pi(l_u+\frac{\beta_u}{\beta_{\phi}}l_{\phi})\coth\frac{\pi l_{\phi}}{\beta_{\phi}}-\beta_u\big]-\frac{c_M}{6}\frac{\epsilon_u}{\epsilon_{\phi}}
\end{equation}
By substituting the following parameters
\begin{equation}\label{beta u and phi}
  \beta_u=\frac{\pi J}{M^{3/2}},~~~\beta_{\phi}=\frac{2\pi}{\sqrt{M}},
\end{equation}
into Eq. (\ref{EE in Poin}) we have
\begin{equation}\label{EE in FSC 2}
  S_{EE}=\frac{c_L}{6}\log\big(\frac{2}{\sqrt{M}\epsilon_{\phi}}\sinh\frac{\sqrt{M} l_{\phi}}{2}\big)+\frac{c_M}{6}\big(\frac{(l_uM+l_{\phi}\sqrt{M}r_c)\coth\frac{\sqrt{M} l_{\phi}}{2}-2r_c}{2\sqrt{M}}\big)-\frac{c_M}{6}\frac{\epsilon_u}{\epsilon_{\phi}}
\end{equation}
By inserting the coordinate extensions as follows \cite{24}
\begin{eqnarray}\label{phi extension in FSC}
  \log\big(\frac{2}{\sqrt{M}\epsilon_{\phi}}\sinh\frac{\sqrt{M} l_{\phi}}{2}\big)&=&\frac{\sqrt{\tilde{M}}}{2}\Delta\tilde{\phi},\\
  \frac{(l_uM+l_{\phi}\sqrt{M}r_c)\coth\frac{\sqrt{M} l_{\phi}}{2}-2r_c}{2\sqrt{M}}-\frac{\epsilon_u}{\epsilon_{\phi}} &=& \sqrt{\tilde{M}}\Delta\tilde{u}+\tilde{r}_c\Delta\tilde{\phi} \label{u extension in FSC}
\end{eqnarray}
in the Eq.(\ref{EE in FSC 2}) we obtain following expression  for the  entanglement entropy
\begin{equation}\label{EE in FSC 3}
  S_{EE}=\frac{c_M}{12}( \sqrt{\tilde{M}}\Delta\tilde{u}+\tilde{r}_c\Delta\tilde{\phi})+\frac{c_L}{12}\sqrt{\tilde{M}}\Delta\tilde{\phi}
\end{equation}

By substituting the central charges Eq.~({\ref{cent ch.}) into the above equation we obtain an expression similar to the entanglement entropy for the zero temperature
BMSFT on the plane where we presented in Eq.~(\ref{EE in zero BMSFT}). \footnote{Note that $\tilde{r}_{c}=\frac{\tilde{J}}{2\sqrt{\tilde{M}}}$}. The last case of the BMSFT is considered on a cylinder with zero temperature. The cylinder identification is as follows
\begin{equation}\label{phi interval}
  \phi \sim \phi+2\pi
\end{equation}

By the similar steps as former cases, the parameters of the Cardy formula in this case are obtained as follows  \cite{23}
\begin{eqnarray} \label{rindler003}
  a &=& \tilde{\beta}_{\phi},~~~~ \bar{a}=\tilde{\beta}_u \nonumber \\
  2\pi b &=& - \frac{\tilde{\beta}_{\phi}}{\pi}\zeta,~~~~ 2\pi\bar{b}=-\frac{\tilde{\beta}_{u}}{\pi}\zeta+\frac{\tilde{\beta_{\phi}}l_u\cot(l_{\phi}/2)}{2\pi}-
  \frac{\tilde{\beta_{\phi}}\epsilon_u}{\pi\epsilon_{\phi}}.
\end{eqnarray}

Substituting these parameters in the Cardy formula (\ref{Cardy formula}), the entanglement entropy with zero temperature on a cylinder is obtained as follows
\begin{equation}\label{EE in global Min}
  S_{EE}=\frac{c_L}{6}\log\big(\frac{2}{\epsilon_{\phi}}\sin\frac{l_{\phi}}{2}\big)+
  \frac{c_M}{12}(l_u\cot\frac{l_{\phi}}{2}-2\frac{\epsilon_u}{\epsilon_{\phi}})
\end{equation}

The coordinate extensions in this case are as follows
\begin{eqnarray}\label{coor extension in global Min}
 \frac{1}{2} \sqrt{\tilde{M}}\Delta\tilde{\phi} &=& \log\big(\frac{2}{\epsilon_{\phi}}\sin\frac{l_{\phi}}{2}\big) \\
   \sqrt{\tilde{M}}\Delta\tilde{u}+\tilde{r}_c\Delta\tilde{\phi} &=& l_u\cot\frac{l_{\phi}}{2}-2\frac{\epsilon_u}{\epsilon_{\phi}}
\end{eqnarray}

Substituting this coordinate extensions in the entropy Eq.~(\ref{EE in global Min}), we obtain the same result for the entanglemet entropy as Eq.~(\ref{EE in FSC 3}).

\section{Holographic entanglement entropy in gravity side}
In the field theory side, a Rindler transformation is used and the BMSFT is mapped to the $\widetilde{BMSFT}$, thus the entanglement entropy of the former is obtained from the thermal entropy of the latter. In the gravity side, we can find a coordinate transformation that transform the spacetime Eq.~(\ref{general met 2}) into a $\widetilde{FSC}$, and consider the latter as a dual with the $\widetilde{BMSFT}$ on the boundary. An important condition is that the new coordinates satisfy the Bondi gauge conditions as follows
\begin{eqnarray}
  g_{\tilde{u},\tilde{u}} &=& \tilde{M},~~~~g_{\tilde{u},\tilde{\phi}}=\tilde{J}\nonumber  \\
  g_{\tilde{\phi},\tilde{\phi}} &=& \tilde{r}^2,~~~~g_{\tilde{r},\tilde{r}}=0\nonumber \\
  g_{\tilde{r},\tilde{\phi}} &=& 0,~~~~g_{\tilde{u},\tilde{r}}=g_{\tilde{u},\tilde{r}}(\tilde{r})
\end{eqnarray}

In Poincare coordinates in the limit $r\to 0$, the transformations between Poincare coordinates and the $\widetilde{FSC}$ are as follows \cite{24}
\begin{eqnarray}
  \tilde{\phi} &=& \frac{2}{\sqrt{\tilde{M}}}\tanh^{-1}\frac{2\phi}{l_{\phi}} \\
  \tilde{u} &=& \frac{4}{\sqrt{\tilde{M}}}\frac{(ul_{\phi}-l_u\phi)}{(l^2_{\phi}-4\phi^2)}-\frac{\tilde{r}_c}{\sqrt{\tilde{M}}}\tilde{\phi}
\end{eqnarray}

With these transformations and the regularized interval Eq.~(\ref{reg. int}), we obtain the extensions in $u$ and $\phi$ directions as follows
\begin{eqnarray}
  \Delta\tilde{u} &=& \frac{2}{\sqrt{\tilde{M}}}(\frac{l_u}{l_{\phi}}-\frac{\epsilon_u}{\epsilon_{\phi}})\frac{\tilde{J}}{2\tilde{M}}\Delta\tilde{\phi} \\
  \Delta\tilde{\phi} &=& \frac{2}{\sqrt{\tilde{M}}}\log\frac{l_{\phi}}{\epsilon_{\phi}}
\end{eqnarray}
Generalized Minimal Massive Gravity (GMMG) was introduced in \cite{11}, providing a new example of a
theory that avoids the bulk-boundary clash and therefore, the theory possesses both, positive energy excitations around the maximally $AdS_{3}$ vacuum
as well as a positive central charge in the dual CFT.
The lagrangian of GMMG model is as follows \cite{11}
\begin{equation}\label{L GMMG}
  L_{GMMG}=L_{GMG}-\frac{\alpha}{2}e.h\times h
\end{equation}
where
\begin{equation}\label{L GMG}
  L_{GMG}=L_{TMG}-\frac{1}{m^2}\big(f.R+\frac{1}{2}e.f\times f\big)
\end{equation}
and
\begin{equation}\label{L TMG}
  L_{TMG}=-\sigma e.R+\frac{\Lambda_0}{6}e.e\times e+h.T(\omega)+\frac{1}{2\mu}\big(\omega.d\omega+
  \frac{1}{3}\omega.\omega\times\omega\big)
\end{equation}

Here, $L_{TMG}$ is the lagrangian of the topological massive gravity and the last term in it is a Lorentz Chern-Simons term. The $\Lambda_0$ is a cosmological parameter and $\sigma$ is a sign. $\alpha $ is a dimensionless parameter, $e$ is a dreibein, $h$
is the auxiliary field, $ \omega $ is a dualized spin-connection, $T(\omega)$ and
$R(\omega)$ are a Lorentz covariant torsion and a curvature 2-form respectively. We need the central charges in the flat limit, which we used in the previous section. The GMMG model has a rotating BTZ black hole solutions as \cite{28}
\begin{equation}\label{BTZ}
  ds^2=-\frac{(r^2-r_+^2)(r^2-r_-^2)}{l^2r^2}dt^2+\frac{l^2r^2}{(r^2-r_+^2)(r^2-r_-^2)}dr^2+r^2(d\phi-\frac{r_+r_-}{lr^2}dt)^2
\end{equation}
where $r_+$ and $r_-$ are outer and inner horizons respectively. The energy and the angular momentum of the BTZ black hole are as follows \cite{28}}
\begin{eqnarray}\label{E , J 1}
  E &=& (\sigma+\frac{\gamma}{2\mu^2 l^2}+\frac{s}{2m^2l^2})\frac{r_+^2+r_-^2}{l^2}-\frac{2r_+r_-}{\mu l^3}\nonumber \\
  J &=& (\sigma+\frac{\gamma}{2\mu^2 l^2}+\frac{s}{2m^2l^2})\frac{2r_+r_-}{l}-\frac{r_+^2+r_-^2}{\mu l^2}
\end{eqnarray}

If one writes the BTZ black hole in terms of the mass parameter $M$ and the angular momentum parameter $a$, the energy and the angular momentum can be written as follows
\begin{eqnarray}\label{E , J 2}
  E &=& (\sigma+\frac{\gamma}{2\mu^2 l^2}+\frac{s}{2m^2l^2})M-\frac{a}{\mu l^2}\nonumber \\
  J &=& (\sigma+\frac{\gamma}{2\mu^2 l^2}+\frac{s}{2m^2l^2})a-\frac{M}{\mu}
\end{eqnarray}

By comparing $E$ and $J$ in Eq.~(\ref{E , J 2}) with corresponding quantities in Eq.~(\ref{E , J 1}), we find the following result for the horizons
\begin{eqnarray}
  r_+r_- &=& \frac{al}{2} \\
  r_+^2+r_-^2 &=& Ml^2
\end{eqnarray}
In the flat limit $r\to \infty$, we have the result as follows
\begin{eqnarray}\label{in and out h}
  r_+ &\to & l\sqrt{M}\nonumber \\
  r_- &\to & \frac{a}{2\sqrt{M}}\equiv r_c
\end{eqnarray}

The general form of the metric is also written using the horizons and in the flat limit as follows
\begin{eqnarray}
  ds^2_{outer} &=& (r_+d\phi+\frac{r_-dt}{l})^2~~\to~~(\sqrt{M}ld\phi+r_c\frac{dt}{l})^2 \\ \label{out line}
  ds^2_{inner} &=& (r_-d\phi+\frac{r_+dt}{l})^2~~\to~~(r_cd\phi+\sqrt{M}dt)^2  \label{inn line}
\end{eqnarray}
which we can use it to calculate the entropy. The thermal entropy of inner horizon of the BTZ black hole is as follows \cite{31',32',33'}
\begin{equation}\label{gen entropy}
  S_{inner}=\frac{l_{inner~horizon}}{4G}+\frac{l_{outer~horizon}}{4G\mu l}
\end{equation}

We do not write the entropy of the outer horizon because we want to do the calculations in the flat limit. So we have
\begin{equation}\label{entropy gravity}
  S_{inner}|_{l\to \infty}=\frac{\sqrt{M}\Delta t+r_c\Delta\phi}{4G}+\frac{\sqrt{M}l\Delta\phi}{4G\mu l}\equiv S_{BH}+S_{CS}
\end{equation}
where the first term is a Bekenestein-Hawking entropy and the second term is related to Chern-Simons contributions in the Lagrangian. By comparing the entanglement entropy Eq.~(\ref{EE in zero BMSFT}) of BMSFT with zero temperature on the plane with Eq.~(\ref{entropy gravity}), the CS correction to the entanglement entropy in Poincare patch is obtained as follows
\begin{equation}\label{CS in Poin}
  S_{CS}=-\frac{1}{2\mu G}\log\frac{l_{\phi}}{\epsilon_{\phi}}
\end{equation}

We can find the holographic entanglement entropy for other cases of the duality. To this end,  we first consider the FSC case. The gravity theory in this case has a holographic dual on the boundary which is a BMS-field theory with finite temperature. The entanglement entropy of the field theory side is obtained in Eq. (\ref{EE in FSC 3}) for a BMSFT with finite temperature. In this case the coordinate extension Eqs. (\ref{phi extension in FSC}) and (\ref{u extension in FSC}) are used to find the entropy in $\tilde{u}$ and $\tilde{\phi}$ coordinates. Substituting the central charges of GMMG in flat limit Eq.~(\ref{cent ch.}) into Eq. (\ref{EE in FSC 3}), we obtain the entropy as follows
\begin{equation}\label{Final EE ins FSC}
  S_{EE}=-\frac{1}{4G}(\bar{\sigma}+\frac{\alpha H}{\mu}+\frac{F}{m^2})
  (\sqrt{\tilde{M}}\Delta\tilde{u}+\tilde{r}_c\Delta\tilde{\phi})-\frac{1}{4\mu G}\sqrt{\tilde{M}}\Delta\tilde{\phi}
\end{equation}

Corresponding its holographic dual in gravity side Eq.~(\ref{entropy gravity}), we obtain the CS correction to the entropy in FSC as follows
\begin{eqnarray}\label{CS in FSCC}
  S_{CS} &=& -\frac{1}{4\mu G}\sqrt{\tilde{M}}\Delta\tilde{\phi}\nonumber\\
         &=& -\frac{1}{2\mu G}\log\big(\frac{2}{\sqrt{M}\epsilon_{\phi}}\sinh\frac{\sqrt{M} l_{\phi}}{2}\big)\
\end{eqnarray}
where we have used Eq.(\ref{phi extension in FSC}) to obtain the 2nd line. The holographic dual of GMMG model in the global Minkowski space is a BMSFT with zero temperature on a cylinder. The entanglement entropy of BMSFT with zero temperature on a cylinder is obtained in Eq. (\ref{EE in global Min}). Substituting the coordinate extensions Eq. (\ref{coor extension in global Min}), we find the following result for entanglement entropy in the field theory side

\begin{equation}\label{EE in global Min 3}
  S_{EE}=\frac{c_M}{12}
  (\sqrt{\tilde{M}}\Delta\tilde{u}+\tilde{r}_c\Delta\tilde{\phi})+\frac{c_L}{12}\sqrt{\tilde{M}}\Delta\tilde{\phi}
\end{equation}

Substituting the central charges Eq.~(\ref{cent ch.}) into this equation, the entanglement entropy obtained as follows
\begin{equation}\label{EE in global Min 3}
  S_{EE}=-\frac{1}{4G}(\bar{\sigma}+\frac{\alpha H}{\mu}+\frac{F}{m^2})
  (\sqrt{\tilde{M}}\Delta\tilde{u}+\tilde{r}_c\Delta\tilde{\phi})-\frac{1}{4\mu G}\sqrt{\tilde{M}}\Delta\tilde{\phi}
\end{equation}

This result is the same as Eq.~(\ref{Final EE ins FSC}) found here again. If we compare this result with the entropy in the gravitational side, we get the CS-correction to the entropy in global Minkowski space as follows
\begin{eqnarray}\label{CS in global Min}
  S_{CS} &=& -\frac{1}{4\mu G}\sqrt{\tilde{M}}\Delta\tilde{\phi}\nonumber\\
   &=& -\frac{1}{2\mu G}\log\big(\frac{2}{\epsilon_{\phi}}\sin\frac{l_{\phi}}{2}\big)
\end{eqnarray}
In this section we considered the GMMG model and obtained the holographic entanglement entropy in three different asymptotically flat solutions of this model: Poincare patch, global Minkowski, and FSC solutions. Our results are coincide exactly with field theory calculations which have been done in previous section. The asymptotically symmetry algebra of GMMG model in contrast with Einstein gravity and similar with TMG has two non vanishing central charges $C_{L}$ and $C_{M}$. By our this study, we have taken into account the effect of non vanishing central charge  $C_{L}$ in holographic entanglement entropy. Equations (\ref{CS in Poin}),(\ref{CS in FSCC}),(\ref{CS in global Min}), are effects of non vanishing central charge $C_{L}$ on holographic entanglement entropy respectively in Poincare patch, FSC, and global Minkowski solution of GMMG model in flat limit. As one can see from these equations, this effect appears as a Chern-Simons correction to the thermal entropy.
\section{Conclusion}
We know that the pure Einstein-Hilbert gravity in three dimensions (in the presence of negative cosmological constat also) exhibits no
propagating physical degrees of freedom \cite{30,31}. Adding the gravitational Chern-Simons term produces a propagating
massive graviton \cite{32}. The resulting theory is called topologically massive gravity (TMG). Unfortunately TMG has a bulk-boundary unitarity conflict. But, as mentioned in introduction, fortunately GMMG avoids the aforementioned \emph{bulk-boundary unitarity clash}.
The calculation of the GMMG action to quadratic order about the $AdS_3$ space shows that the theory is free of negative energy
bulk modes. So this model is a viable candidate for the semi-classical limit of a unitary quantum 3D massive gravity. Our motivation in this paper is that to give another evidence on the duality between GMMG in the bulk and a quantum field theory in the boundary. Here instead of $AdS_3$ space we consider asymptotically flat space, and study the holographic duality in the flat limit of GMMG.\\
The authors of \cite{24} have obtained the holographic entanglement entropy of a single interval in the dual BMSFT located at the null infinity of the
$3-$dimensional bulk asymptotically flat spacetime in the framework of Einstein gravity and TMG. In the present work we have extended such study to the GMMG model. In the another term to give an interesting and important evidence for this duality, we calculate entanglement entropy at the bulk and at the boundary. To calculate entanglement entropy on the boundary, we consider a BMSFT on the boundary, and then by placing its manifold on the thermal circle Eq.(\ref{thermal circle}), we calculate the entropy on the interval of the coordinate extensions using the Cardy formula Eq.~(\ref{Cardy formula}). To calculate the coordinate extensions, the Rindler transformation is used and these extensions are considered in three cases with zero temperature on the plane and cylinder and finite temperature. What characterizes the GMMG model are the central charges Eq.~(\ref{cent ch.}) that we substitute in the entropy formula and get the result Eq.~(\ref{EE in zero BMSFT}).

Using the expressions of the energy and angular momnetum Eqs.~(\ref{E , J 1}) and ~(\ref{E , J 2}), for BTZ black hole solution of GMMG, we could obtain the outer and inner horizons Eq.~(\ref{in and out h}) in the flat limit, where we need to obtain entanglement entropy in the bulk.
We have obtained some corrections to Beckenstein-Hawking entropy, which are not only due to the presence of Chern-Simons term in the Lagrangian of the model, but from higher order curvature terms. The contribution of Chern-Simons term in the entanglement entropy of bulk is exactly the same has been obtained by Jiang et.al \cite{24} for topological massive gravity. It is interesting that the contribution of another terms of the Lagrangian of GMMG model appear as a coefficient of usual Beckenstein-Hawking term in the entanglement entropy.

\end{document}